\documentclass[english]{article}
\usepackage[T1]{fontenc}
\usepackage[latin9]{inputenc}
\usepackage{calc}
\usepackage{amsmath}
\usepackage{amssymb}
\usepackage{graphicx}

\makeatletter
\usepackage{amsfonts,amssymb}
\usepackage{latexsym}
\usepackage{epsfig}
\usepackage{cite}
\usepackage{graphicx}
\usepackage[colorlinks,linkcolor=blue]{hyperref}
\usepackage{tikz}

\newcommand{\be}{\begin{equation}}
\newcommand{\ee}{\end{equation}}

\makeatother

\usepackage{babel}
\begin{document}
{}~ \hfill\vbox{\hbox{hep-th/yymmnnn}}\break
\vskip 3.0cm
\centerline{\Large \bf  Transition between Schwarzschild black hole and string black hole}

\vspace*{10.0ex}

\vspace*{10.0ex}
\centerline{\large Shuxuan Ying}
\vspace*{7.0ex}
\vspace*{4.0ex}

\centerline{\large \it Department of Physics}
\centerline{\large \it Chongqing University}
\centerline{\large \it Chongqing, 401331, China} \vspace*{1.0ex}

\vspace*{4.0ex}
\centerline{\large ysxuan@cqu.edu.cn}

\vspace*{4.0ex}

\centerline{\bf Abstract} \bigskip \smallskip

In this paper, we aim to study the quantum transition between a Schwarzschild black hole and a string black hole in the large $D$ limit. Classically, such a transition between these two distinct black hole geometries is forbidden. The only feasible discussion is centered on how a black hole evaporates, loses mass, and transitions into highly excited fundamental strings. Building upon our previous work on T-duality between the Schwarzschild and string black holes, we reduce the problem to two dimensions, where the corresponding Wheeler-De Witt equation can be derived. Using this equation, we identify the two black hole geometries as distinct wave function states. This allows us to easily compute the transition probability between these two geometries, driven by the string coupling.

\vfill \eject
\baselineskip=16pt
\vspace*{10.0ex}
\tableofcontents

\section{Introduction}

The evolution of a Schwarzschild black hole into highly excited fundamental
strings through mass loss during evaporation is an intriguing problem
\cite{Bowick:1985af,Susskind:1993ws,Horowitz:1996nw,Horowitz:1997jc},
as it bridges two distinct regimes: the black hole interior and string
microstates \cite{Chen:2021emg}. When the black hole\textquoteright s
mass is sufficiently large, the highly excited fundamental strings
form a single long string, whose self-gravitation gives rise to a
string star. The string star continues to evaporate, eventually transitioning
into free strings. Traditionally, this process has been studied using
the Horowitz-Polchinski effective field theory \cite{Horowitz:1997jc}.
Relevant discussions are provided in the refs \cite{Chen:2021dsw,Ceplak:2023afb,Chu:2024ggi}.
In this paper, we approach the problem from a new perspective. We
do not focus on how the Schwarzschild black hole directly evolves
into fundamental strings; instead, we investigate its intermediate
step, specifically how the Schwarzschild black hole transitions into
the string black hole \footnote{Here, a string black hole refers to a black hole solution of the low-energy effective action in string theory, featuring a nontrivial dilaton field.}.

\begin{figure}[h]
\begin{centering}
\includegraphics[scale=0.8]{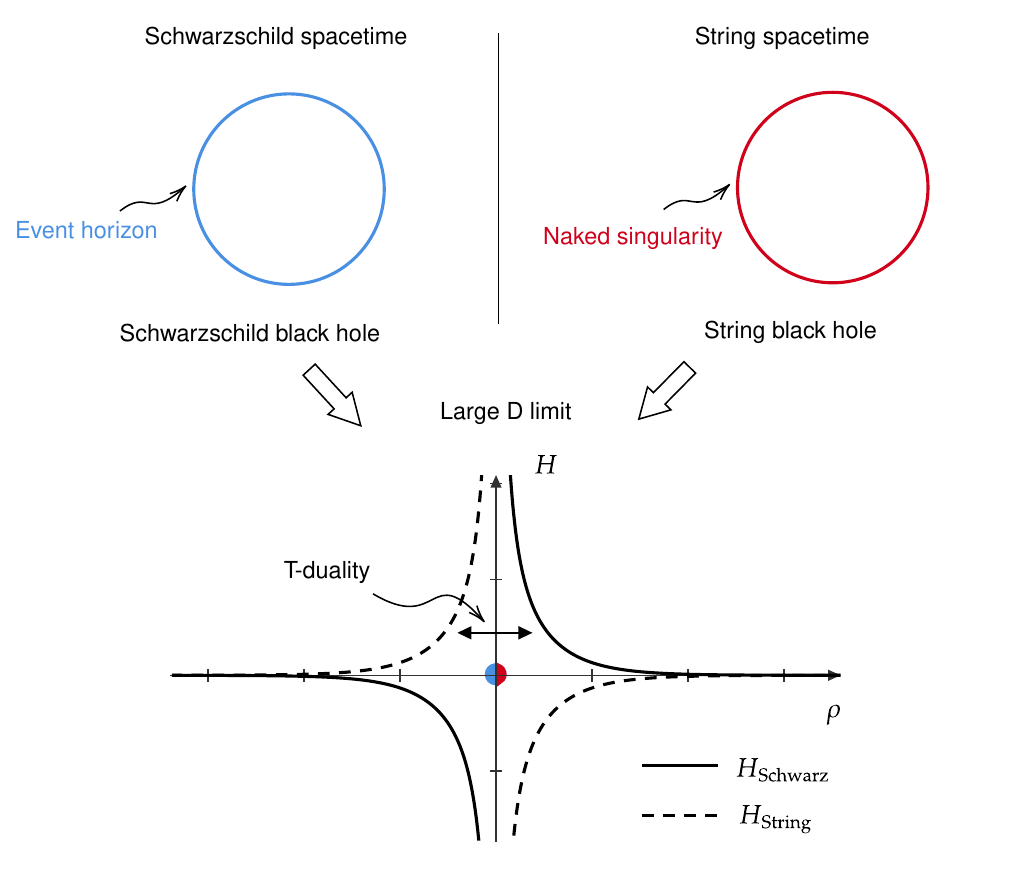}
\par\end{centering}
\caption{\label{fig:idea}This figure depicts our strategy for connecting two
distinct black holes---one from Einstein's gravity and the other
from string theory. In the large $D$ limit, the near horizon geometries
of Schwarzschild and string black holes reduce to two T-dual, two-dimensional
black strings that cover different regions of the two-dimensional
spacetime. In this figure, $H$ represents the Hubble-like parameter
for the background, and $\rho$ is a spacelike coordinate. Detailed
definitions and further explanations can be found in the rest of the
paper.}
\end{figure}

Classically, the transition from a Schwarzschild black hole to a string
black hole is forbidden, as they represent solutions to different
actions---the Einstein-Hilbert action and the low-energy effective
action of string theory, respectively. However, quantum dynamics allows
transitions between these two distinct geometric configurations. Our
goal is to describe this evolution using a quantum framework. Although
the Schwarzschild and string black holes arise from different actions,
their connection can be established through key observations:
\begin{enumerate}
\item The low-energy effective action admits two kinds of black hole solutions.
The first is the string black hole, with a non-trivial dilaton and
a naked singularity but no event horizon. The second is the Schwarzschild
black hole, with a constant or vanishing dilaton. These two solutions
are T-dual to each other \cite{Tseytlin:1991wr,Ginsparg:1992af,Kar:1998rv,Exirifard:2004ey};
the dilaton vanishes in the T-dual transformation, linking the two
black hole solutions.
\item In the framework of large $D$ gravity \cite{Emparan:2013xia}, near-horizon
geometries of both solutions reduce to two-dimensional black strings,
whose relationship is described by the well-known scale-factor duality
of two-dimensional low energy effective theory \cite{Ying:2024wrv}.
This simplification reduces the complex problem to a two-dimensional
framework.
\end{enumerate}
Crucially, the two-dimensional reduced solutions from Schwarzschild
and string black holes share the same low-energy effective action
and cover different regions of the two-dimensional spacetime, as illustrated
in Figure (\ref{fig:idea}). This allows us to employ the Wheeler-De
Witt (WDW) approach to quantize the action. Using the WDW equation,
the spacetime can be represented by a wave function evolving in superspace,
where each point corresponds to a specific geometric configuration.
Quantum dynamics can then facilitate transitions between these configurations,
enabling a transition from a Schwarzschild black hole to a string
black hole. If this transition is possible in the simplified two-dimensional
model, it suggests that such a process is also feasible in higher-dimensional
spacetimes, as the dimensionality does not affect the underlying quantum
mechanics.

In this paper, we first begin with T-dual black hole solutions of
low-energy effective action of string theory. These two black hole
solutions relates to the Schwarzschild and string black holes:

\begin{eqnarray}
ds_{\mathrm{Schwarz}}^{2}=-\left(1-\left(\frac{r_{0}}{r}\right)^{n}\right)dt^{2}+\frac{dr^{2}}{\left(1-\left(\frac{r_{0}}{r}\right)^{n}\right)}+r^{2}d\Omega_{n+1}^{2}, &  & \phi_{\mathrm{Schwarz}}\left(r\right)=0,\nonumber \\
ds_{\mathrm{String}}^{2}=-\left(1-\left(\frac{r_{0}}{r}\right)^{n}\right)^{-1}dt^{2}+\frac{dr^{2}}{\left(1-\left(\frac{r_{0}}{r}\right)^{n}\right)}+r^{2}d\Omega_{n+1}^{2}, &  & \phi_{\mathrm{String}}\left(r\right)=-\frac{1}{2}\ln\left(1-\left(\frac{r_{0}}{r}\right)^{n}\right).
\end{eqnarray}

\noindent In the large $D$ limit, these two black hole solutions
reduce to two T-dual black string solutions.

\begin{eqnarray}
ds_{\mathrm{Schwarz}}^{2} & = & \left(\frac{2r_{0}}{n}\right)^{2}\left(-\tanh^{2}\rho d\tau^{2}+d\rho^{2}\right),\qquad\phi\left(\rho\right)=-\frac{1}{2}\ln\cosh^{2}\rho,\nonumber \\
ds_{\mathrm{String}}^{2} & = & \left(\frac{2r_{0}}{n}\right)^{2}\left(-\coth^{2}\rho d\tau^{2}+d\rho^{2}\right),\qquad\tilde{\phi}\left(\rho\right)=-\frac{1}{2}\ln\sinh^{2}\rho.
\end{eqnarray}

\noindent with the following coordinate transformation:

\begin{equation}
\left(\frac{r}{r_{0}}\right)^{n}=\cosh^{2}\rho,
\end{equation}
and $\tilde{\phi}\left(\rho\right)=\phi\left(\rho\right)-\frac{1}{2}\ln\left(-g_{00}\right)$
is called the shifted dilaton \cite{Gasperini:1991ak}. In the near-horizon
region, $r\geq r_{0}$ and $\mathrm{R}\geq1$, since $\cosh\left(\rho\right)\geq1$,
the coordinate $\rho$ can range from $-\infty$ to $+\infty$. Therefore,
we can select a possible combination of solutions where $ds_{\mathrm{String}}^{2}$
describes the region $\rho<0$, and $ds_{\mathrm{Schwarz}}^{2}$ describes
the region $\rho>0$. Together, these two solutions cover the entire
two-dimensional spacetime. As expected, these two solutions are T-dual
to each other in the context of the two-dimensional low-energy effective
action with a nonvanishing cosmological constant. This result is analogous
to string cosmology, allowing us to adopt well-established methods
from that framework to study this problem \cite{Gasperini:1996fn,Gasperini:1996np,Gasperini:1997uh,Gasperini:2007zz,Gasperini:2021eri}.
Starting with the action, we derive the corresponding Hamiltonian
constraint and WDW equation. By solving the WDW equation and imposing
the tunneling boundary conditions, the wave function is uniquely determined.
The physical picture is as follows: we begin with the Schwarzschild
initial state, $\Psi_{+}\left(\beta,\Phi\rightarrow-\infty\right)$,
in the low-energy regime. A part of this wave transmits to the singularity
as $\Psi\left(\beta,\Phi\rightarrow+\infty\right)$, while the remaining
portion, $\Psi_{-}\left(\beta,\Phi\rightarrow-\infty\right)$ reflects
to the string black hole. The transition probability is determined
by the reflection coefficient:

\begin{equation}
P=\frac{\left|\Psi_{-}\left(\beta,\Phi\rightarrow-\infty\right)\right|^{2}}{\left|\Psi_{+}\left(\beta,\Phi\rightarrow-\infty\right)\right|^{2}}=\exp\left(-2k\pi\right),
\end{equation}

\noindent where $k=c\lambda$ relates to the cosmological constant
$4\lambda^{2}$, and $c$ is the constant of integration. This result
demonstrates that the classically forbidden transition between Schwarzschild
and string black holes becomes possible within the framework of quantum
theory, thereby establishing a connection between Einstein's gravity
and string theory at the quantum level. It further implies the possibility
that a black hole with an event horizon could transition into a geometry
with a naked singularity.

This paper is organized as follows: In Section 2, we provide a review
of the higher-dimensional black hole solutions and their T-dual counterparts
in string theory. We then demonstrate the scale-factor duality between
these two black hole solutions in the large $D$ limit. In Section
3, we derive the WDW equation for the two-dimensional low-energy effective
action and compute the transition probability between the Schwarzschild
black hole and the string black hole. Finally, in Section 4, we present
a discussion and conclude the paper. 

\section{Brief review of T-dual large $D$ black holes}

In our previous work \cite{Ying:2024wrv}, we obtained the $D$-dimensional
T-dual black hole solutions from the low-energy effective string theory.
We begin with the low-energy effective action of string theory in
D dimensions, assuming a vanishing Kalb-Ramond field and cosmological
constant:

\begin{equation}
I_{\mathrm{String}}=\frac{1}{16\pi G_{D}}\int d^{D}x\sqrt{-g}e^{-2\phi}\left(R+4\left(\partial\phi\right)^{2}\right),
\end{equation}

\noindent where $\phi$ represents the $D$-dimensional dilaton field.
To extend the four-dimensional black hole solution \cite{Kar:1998rv}
to higher dimensions and its T-dual, we proceed as follows:

\vspace*{2.0ex}

\noindent \textbf{Ordinary solution:}

\noindent The ordinary higher-dimensional black hole solution is given
by:

\begin{equation}
ds_{\mathrm{Schwarz}}^{2}=-\left(1-\left(\frac{2\eta}{r}\right)^{n}\right)^{\frac{m+\sigma}{\eta}}dt^{2}+\left(1-\left(\frac{2\eta}{r}\right)^{n}\right)^{\frac{\sigma-m}{\eta}}dr^{2}+\left(1-\left(\frac{2\eta}{r}\right)^{n}\right)^{1+\frac{\sigma-m}{\eta}}r^{2}d\Omega_{n+1}^{2},\label{eq:D string black hole}
\end{equation}

\noindent where $d\Omega_{n+1}^{2}$ denotes the metric of an $\left(n+1\right)$-dimensional
sphere, and the dilaton solution is:

\begin{equation}
\phi_{\mathrm{Schwarz}}\left(r\right)=\frac{1}{4\eta}\left(\left(n-1\right)\left(\eta-m\right)+\left(n+1\right)\sigma\right)\ln\left(1-\left(\frac{2\eta}{r}\right)^{n}\right),
\end{equation}

\noindent subject to the constraint:

\begin{equation}
-\left(n+1\right)m^{2}-\left(n-3\right)\eta^{2}-\left(n+1\right)\sigma^{2}+2\left(n-1\right)m\left(\eta+\sigma\right)-2\left(n-1\right)\eta\sigma=0.\label{eq:constraint}
\end{equation}

\vspace*{2.0ex}

\noindent \textbf{T-dual solution:}

\noindent Applying the Buscher rules with a vanishing Kalb-Ramond
field along the $g_{00}$ direction,

\begin{equation}
\tilde{g}_{00}=\frac{1}{g_{00}},\qquad\phi_{\mathrm{String}}=\phi_{\mathrm{Schwarz}}-\frac{1}{2}\ln\left(-g_{00}\right),
\end{equation}

\noindent the T-dual black hole solution is:

\begin{equation}
ds_{\mathrm{String}}^{2}=-\left(1-\left(\frac{2\eta}{r}\right)^{n}\right)^{\frac{-m-\sigma}{\eta}}dt^{2}+\left(1-\left(\frac{2\eta}{r}\right)^{n}\right)^{\frac{\sigma-m}{\eta}}dr^{2}+\left(1-\left(\frac{2\eta}{r}\right)^{n}\right)^{1+\frac{\sigma-m}{\eta}}r^{2}d\Omega_{n+1}^{2},\label{eq:D string black hole dual}
\end{equation}

\noindent with the shifted dilaton solution:

\begin{equation}
\phi_{\mathrm{String}}\left(r\right)=\frac{1}{4\eta}\left(\left(n-1\right)\left(\eta+\sigma\right)-\left(n+1\right)m\right)\ln\left(1-\left(\frac{2\eta}{r}\right)^{n}\right).
\end{equation}

\noindent The constraint for the parameters is same as (\ref{eq:constraint}).

\vspace*{2.0ex}

To maintain spherical symmetry in both metrics, we choose identical
values for $m$, $\eta$, and $\sigma$, such that $\sigma=0$ and
$m=\eta$. Consequently, the metrics (\ref{eq:D string black hole})
and (\ref{eq:D string black hole dual}) are given by:

\begin{eqnarray}
ds_{\mathrm{Schwarz}}^{2}=-\left(1-\left(\frac{r_{0}}{r}\right)^{n}\right)dt^{2}+\frac{dr^{2}}{\left(1-\left(\frac{r_{0}}{r}\right)^{n}\right)}+r^{2}d\Omega_{n+1}^{2}, &  & \phi_{\mathrm{Schwarz}}\left(r\right)=0,\label{eq:ordinary metric}\\
ds_{\mathrm{String}}^{2}=-\left(1-\left(\frac{r_{0}}{r}\right)^{n}\right)^{-1}dt^{2}+\frac{dr^{2}}{\left(1-\left(\frac{r_{0}}{r}\right)^{n}\right)}+r^{2}d\Omega_{n+1}^{2}, &  & \phi_{\mathrm{String}}\left(r\right)=-\frac{1}{2}\ln\left(1-\left(\frac{r_{0}}{r}\right)^{n}\right),\label{eq:T dual metric}
\end{eqnarray}

\noindent where $r_{0}=2m$. Here, $ds_{\mathrm{Schwarz}}^{2}$ represents
the Schwarzschild-Tangherlini black hole solution in Einstein's gravity
with a vanishing dilaton, where $r_{0}$ denotes the event horizon.
On the other hand, $ds_{\mathrm{String}}^{2}$ corresponds to the
string black hole solution with a non-trivial dilaton, where $r_{0}$
indicates the curvature singularity.

To study the large $D$ limit in the metric, we can introduce the
coordinate transformation $\mathrm{R}=\left(\frac{r}{r_{0}}\right)^{n}$.
The near-horizon metrics for the black hole solutions (\ref{eq:ordinary metric})
and (\ref{eq:T dual metric}) can be obtained by requiring $\ln\mathrm{R}\ll n$,
which gives:

\begin{eqnarray}
ds_{\mathrm{Schwarz}}^{2} & = & -\frac{\mathrm{R}-1}{\mathrm{R}}dt^{2}+\frac{r_{0}^{2}}{n^{2}}\frac{1}{\mathrm{R}\left(\mathrm{R}-1\right)}d\mathrm{R}^{2}+r_{0}^{2}d\Omega_{n+1}^{2},\nonumber \\
ds_{\mathrm{String}}^{2} & = & -\frac{\mathrm{R}}{\mathrm{R}-1}dt^{2}+\frac{r_{0}^{2}}{n^{2}}\frac{1}{\mathrm{R}\left(\mathrm{R}-1\right)}d\mathrm{R}^{2}+r_{0}^{2}d\Omega_{n+1}^{2}.
\end{eqnarray}

\noindent The detailed calculations for these two results can be found
in the previous work. Further employing the coordinate transformations:

\begin{equation}
\mathrm{R}=\cosh^{2}\rho,\qquad d\tau=\frac{n}{2r_{0}}dt,\label{eq:R and rho}
\end{equation}

\noindent the metrics become:

\begin{eqnarray}
ds_{\mathrm{Schwarz}}^{2} & = & \left(\frac{2r_{0}}{n}\right)^{2}\left(-\tanh^{2}\rho d\tau^{2}+d\rho^{2}\right),\qquad\phi\left(\rho\right)=-\frac{1}{2}\ln\cosh^{2}\rho,\nonumber \\
ds_{\mathrm{String}}^{2} & = & \left(\frac{2r_{0}}{n}\right)^{2}\left(-\coth^{2}\rho d\tau^{2}+d\rho^{2}\right),\qquad\tilde{\phi}\left(\rho\right)=-\frac{1}{2}\ln\sinh^{2}\rho,\label{eq:2D dual metrics}
\end{eqnarray}

\noindent where $\tilde{\phi}\left(\rho\right)$ is the shifted dilaton,
the detailed discussion can be found in our previous work \cite{Ying:2024wrv}:

\begin{equation}
\tilde{\phi}\left(\rho\right)=\phi\left(\rho\right)+\phi_{\mathrm{String}}\left(\rho\right)=-\frac{1}{2}\ln\cosh^{2}\rho-\frac{1}{2}\ln\tanh^{2}\rho=-\frac{1}{2}\ln\sinh^{2}\rho.
\end{equation}

Before further discussion, let us recall the coordinate transformation
(\ref{eq:R and rho}). When $r\geq r_{0}$, $\mathrm{R}\geq1$. Since
$\cosh\left(\rho\right)\geq1$, it implies that the coordinate $\rho$
can range from $-\infty$ to $+\infty$. To make this clearer, let
us introduce the Hubble-like parameter:

\begin{equation}
H\left(\rho\right)\equiv\frac{\partial_{\rho}a\left(\rho\right)}{a\left(\rho\right)},
\end{equation}

\noindent for the matric

\begin{equation}
ds^{2}=d\rho^{2}-a\left(\rho\right)^{2}d\tau^{2}.
\end{equation}

\noindent the Hubble-like parameter determines the scalar curvature
for this specific form of the metric. Considering the dual metrics
(\ref{eq:2D dual metrics}), the corresponding Hubble-like parameters
are given by:

\begin{equation}
H_{\mathrm{Schwarz}}\left(\rho\right)=2\text{csch}\left(2\rho\right),\qquad H_{\mathrm{String}}\left(\rho\right)=-2\text{csch}\left(2\rho\right).
\end{equation}

\begin{figure}[h]
\begin{centering}
\includegraphics[scale=0.8]{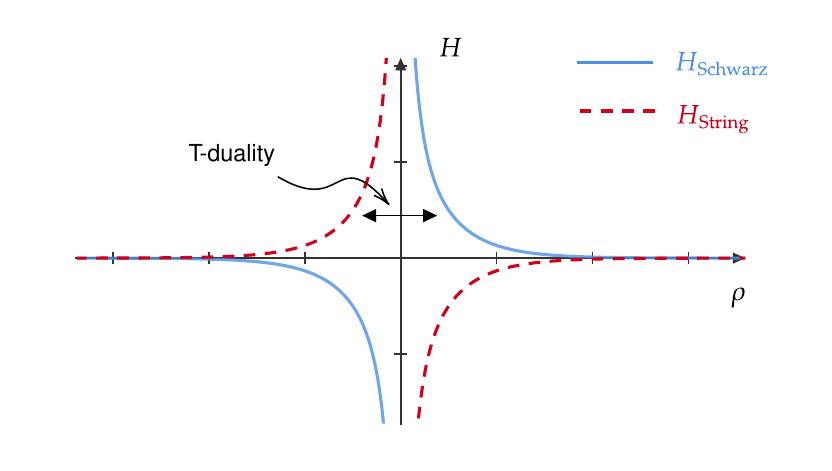}
\par\end{centering}
\caption{\label{fig:Hubble}The Hubble-like parameters for the two-dimensional
black holes reduced from large $D$ limit of Schwarzschild and string
black holes.}
\end{figure}

\noindent Note that $H_{\mathrm{Schwarz}}\left(\rho\right)\longleftrightarrow H_{\mathrm{String}}\left(\rho\right)$
and $\phi\left(\rho\right)\longleftrightarrow\tilde{\phi}\left(\rho\right)$
represent a well-known scale-factor duality \cite{Veneziano:1991ek,Sen:1991zi}.
The Hubble-like parameters can be plotted in the figure (\ref{fig:Hubble}).

\vspace*{2.0ex}

\section{Wheeler--De Witt equation in two spacetimes}

\noindent We have noted in the previous section that the Schwarzschild-Tangherlini
black hole (with a constant or vanishing dilaton $\phi$) and its
T-dual, namely the string black hole, are both solutions to the low-energy
effective action of closed string theory. Since these two black hole
solutions can reduce to the two T-dual string black holes (\ref{eq:2D dual metrics})
in the large $D$ limit, the corresponding $D$-dimensional action
must also reduce to the relevant two-dimensional action simultaneously.
To clarify this, we will consider these two cases separately. 

We first examine the Schwarzschild-Tangherlini black hole with a vanishing
dilaton solution. The action is simply the Einstein-Hilbert action:

\begin{equation}
I_{\mathrm{EH}}=\frac{1}{16\pi G_{D}}\int d^{D}x\sqrt{-g}R.\label{eq:EH action}
\end{equation}

\noindent By performing dimensional reduction on a sphere and introducing
the dilaton field $\phi\left(x^{\mu}\right)$, we write the metric
as:

\begin{equation}
ds^{2}=\underset{2\;\mathrm{dimensions}}{\underbrace{\mathbb{G}_{\mu\nu}\left(x^{\mu}\right)dx^{\mu}dx^{\nu}}}+\underset{n+1\;\mathrm{dimensional\;sphere}}{\underbrace{r_{0}^{2}e^{-4\phi\left(x^{\mu}\right)/\left(n+1\right)}d\Omega_{n+1}^{2}}},\label{eq:metric 2}
\end{equation}

\noindent where the Einstein-Hilbert action becomes:

\begin{equation}
I_{\mathrm{EH}}=\frac{\Omega_{n+1}r_{0}^{n+1}}{16\pi G_{D}}\int d^{2}x\sqrt{-\mathbb{G}}e^{-2\phi}\left(\mathbb{R}+\frac{4n}{n+1}\left(\partial\phi\right)^{2}+\frac{n\left(n+1\right)}{r_{0}^{2}}e^{-4\phi/\left(n+1\right)}\right),
\end{equation}

\noindent where $\mathbb{R}$ is the Ricci scalar of the two-dimensional
metric $\mathbb{G}_{\mu\nu}$, $\phi$ is the two-dimensional dilaton,
and the volume of the unit sphere is given by $\Omega_{n+1}=2\pi^{\frac{n+2}{2}}/\Gamma\left(\frac{n+2}{2}\right)$.
In the large $n$ limit ($n\rightarrow\infty$), the action reduces
to the two-dimensional string effective action. This action possesses
an $SU\left(2\right)_{k}/U\left(1\right)$ symmetry:

\begin{equation}
I_{\mathrm{EH}}^{2D}=\frac{1}{16\pi G_{2}}\int d^{2}x\sqrt{-\mathbb{G}}e^{-2\phi}\left(\mathbb{R}+4\left(\partial\phi\right)^{2}+4\lambda^{2}\right),\label{eq:large D 2D action}
\end{equation}

\noindent where $G_{2}=\underset{n\rightarrow\infty}{\lim}\frac{G_{D}}{\Omega_{n+1}r_{0}^{n+1}}$
and $\lambda=\frac{n}{2r_{0}}$. 

On the other hand, recall the T-dual action for the Einstein-Hilbert
action, which is the low-energy effective action with a non-vanishing
dilaton $\phi_{\mathrm{String}}$:

\begin{equation}
I_{\mathrm{String}}=\frac{1}{16\pi G_{D}}\int d^{D}x\sqrt{-g}e^{-2\phi_{\mathrm{String}}}\left(R+4\left(\partial\phi_{\mathrm{String}}\right)^{2}\right).\label{eq:T-dual action}
\end{equation}

\noindent On the other hand, recall the T-dual action for the Einstein-Hilbert
action, which is the low-energy effective action with a non-vanishing
dilaton

\begin{equation}
ds^{2}=\underset{2\;\mathrm{dimensions}}{\underbrace{\mathbb{G}_{\mu\nu}dx^{\mu}dx^{\nu}}}+\underset{n+1\;\mathrm{dimensional\;sphere}}{\underbrace{r_{0}^{2}e^{-4\phi\left(x\right)/\left(n+1\right)}d\Omega_{n+1}^{2}}},
\end{equation}

\noindent the action (\ref{eq:T-dual action}) becomes:

\begin{equation}
I_{\mathrm{String}}=\frac{\Omega_{n+1}r_{0}^{n+1}}{16\pi G_{D}}\int d^{2}x\sqrt{-\mathbb{G}}e^{-2\left(\phi_{\mathrm{String}}+\phi\right)}\left[\mathbb{R}+4\left(\partial\phi_{\mathrm{String}}\right)^{2}+8\partial\phi\partial\phi_{\mathrm{String}}+\frac{4n}{n+1}\left(\partial\phi\right)^{2}+\frac{n\left(n+1\right)}{r_{0}^{2}}e^{\frac{4\phi}{n+1}}\right],
\end{equation}

\noindent where $\mathbb{R}$ is the Ricci scalar of the two-dimensional
metric $\mathbb{G}_{\mu\nu}$. In the limit $n\rightarrow\infty$,
the action reduces to the two-dimensional low-energy effective action:

\begin{equation}
I_{\mathrm{String}}^{2D}=\frac{1}{16\pi G_{2}}\int d^{2}x\sqrt{-\mathbb{G}}e^{-2\tilde{\phi}}\left(\mathbb{R}+4\left(\partial\tilde{\phi}\right)^{2}+4\lambda^{2}\right),\label{eq:large D 2D dual action}
\end{equation}

\noindent where $\tilde{\phi}\left(\rho\right)\equiv\phi\left(\rho\right)+\phi_{\mathrm{String}}\left(\rho\right)$.
We also use the relations $G_{2}=\underset{n\rightarrow\infty}{\lim}\frac{G_{D}}{\Omega_{n+1}r_{0}^{n+1}}$
and $\lambda=\frac{n}{2r_{0}}$. As expected, the Einstein-Hilbert
action (\ref{eq:EH action}) and low-energy effective action (\ref{eq:T-dual action})
both reduce to the same two-dimensional low-energy effective action
in the large $D$ limit. In the following sections, we write this
action as:

\begin{equation}
I_{\mathrm{String}}^{2D}=\frac{1}{16\pi G_{2}}\int d^{2}x\sqrt{-\mathbb{G}}e^{-2\phi}\left(\mathbb{R}+4\left(\partial\phi\right)^{2}+4\lambda^{2}\right).\label{eq:2D action}
\end{equation}

\noindent The previous two dual black holes (\ref{eq:2D dual metrics}):

\begin{eqnarray}
ds_{\mathrm{Schwarz}}^{2} & = & \left(\frac{2r_{0}}{n}\right)^{2}\left(-\tanh^{2}\rho d\tau^{2}+d\rho^{2}\right),\qquad\phi\left(\rho\right)=-\frac{1}{2}\ln\cosh^{2}\rho,\nonumber \\
ds_{\mathrm{String}}^{2} & = & \left(\frac{2r_{0}}{n}\right)^{2}\left(-\coth^{2}\rho d\tau^{2}+d\rho^{2}\right),\qquad\tilde{\phi}\left(\rho\right)=-\frac{1}{2}\ln\sinh^{2}\rho,
\end{eqnarray}

\noindent are solutions of this action (\ref{eq:2D action}), where
$\tilde{\phi}\left(\rho\right)=\phi\left(\rho\right)+\phi_{\mathrm{String}}\left(\rho\right)$.
To proceed further and obtain the corresponding WDW equation, we utilize
the following coordinate transformation:

\begin{equation}
\frac{2r_{0}}{n}d\tau=d\tau,\qquad\frac{2r_{0}}{n}d\rho=d\rho.
\end{equation}

\noindent Therefore, the solution can be rewritten as:

\begin{eqnarray}
ds_{\mathrm{Schwarz}}^{2} & = & d\rho^{2}-a\left(\rho\right)^{2}d\tau^{2}\equiv d\rho^{2}-\tanh^{2}\left(\lambda\rho\right)d\tau^{2},\qquad\phi\left(\rho\right)=-\frac{1}{2}\ln\cosh^{2}\left(\lambda\rho\right),\nonumber \\
ds_{\mathrm{String}}^{2} & = & d\rho^{2}-a\left(\rho\right)^{-2}d\tau^{2}\equiv d\rho^{2}-\coth^{2}\left(\lambda\rho\right)d\tau^{2},\qquad\tilde{\phi}\left(\rho\right)=-\frac{1}{2}\ln\sinh^{2}\left(\lambda\rho\right),\label{eq:2D final solution}
\end{eqnarray}

\noindent Now, considering the combination of the solutions, we focus
on the following configuration:

\noindent\fbox{\begin{minipage}[t]{1\columnwidth - 2\fboxsep - 2\fboxrule}%
\begin{eqnarray}
\rho<0, &  & a\left(-\rho\right)=\coth\left(-\lambda\rho\right),\qquad\tilde{\phi}\left(-\rho\right)=-\frac{1}{2}\ln\sinh^{2}\left(-\lambda\rho\right),\qquad\mathrm{String\;spacetime},\nonumber \\
\rho>0, &  & a\left(\rho\right)=\tanh\left(\lambda\rho\right),\qquad\phi\left(\rho\right)=-\frac{1}{2}\ln\cosh^{2}\left(\lambda\rho\right),\qquad\mathrm{Schwarzschild\;spacetime}.\label{eq:configuration}
\end{eqnarray}
\end{minipage}}

\vspace{5mm}

\noindent Therefore, the two geometries can be stitched together in
the $\left(\tau,\rho\right)$ plane, allowing us to study their quantum
properties. Next, it is easy to check that the action (\ref{eq:2D action})
is invariant under the following T-dual transformations, also known
as scale-factor duality:

\begin{equation}
a\left(\rho\right)\longleftrightarrow a\left(\rho\right)^{-1},\qquad H\left(\rho\right)\longleftrightarrow-H\left(\rho\right),\qquad\phi\left(\rho\right)\longleftrightarrow\tilde{\phi}\left(\rho\right).
\end{equation}

Now, our goal is to obtain the WDW equatiton from the action (\ref{eq:2D action}).
In quantum mechanics, a wave function controlled by the Schr$\ddot{\mathrm{o}}$dinger
equation is used to describe a particle in Hilbert space. Similarly,
if we treat the whole spacetime as a particle, we can mimic quantum
mechanics to use a wave function to study the quantum dynamics of
the spacetime manifold. These quantum effects allow for transitions
between different geometries that are forbidden in classical theory.
In this analogy, the Hilbert space is replaced by the superspace,
and the wave function is referred to as the WDW equation. In other
words, spacetime can be viewed as a particle state in superspace.
For our case, the WDW equation equation in the two-dimensional superspace
$\left\{ \beta\left(\rho\right),\Phi\left(\rho\right)\right\} $,
called minisuperspace, which is defined by:

\begin{equation}
\beta\left(\rho\right)=\ln a\left(\rho\right),\qquad\Phi\left(\rho\right)=2\phi\left(\rho\right)-\beta\left(\rho\right)-\ln\int d\rho\left(8\pi G_{2}\right),\label{eq:minisuperspace}
\end{equation}

\noindent where each point $a\left(\rho\right)$, $\Phi\left(\rho\right)$
of minisuperspace is the classical solution of (\ref{eq:2D final solution}),
and we assume that $\int d\rho<\infty$. To derive the WDW equation
using the variables $\beta\left(\rho\right)$ and $\Phi\left(\rho\right)$
in the minisuperspace, we must recover the temporal-like gauge $n\left(\rho\right)$.
Thus, the ansatz for the metric is given by:

\begin{equation}
ds^{2}=n\left(\rho\right)^{2}d\rho^{2}-a\left(\rho\right)^{2}d\tau^{2}.
\end{equation}

\noindent The action (\ref{eq:2D action}) takes the form:

\begin{equation}
I_{\mathrm{String}}^{2D}=\int d\rho e^{-\Phi}\frac{1}{2n}\left[\Phi^{\prime2}-\beta^{\prime2}+n^{2}\left(4\lambda^{2}\right)\right],
\end{equation}

\noindent where the prime denotes the derivative with respect to $\rho$.
Note that $\Phi$ is also referred to as the $O\left(d,d\right)$
dilaton, $\beta^{\prime}=H$ is the Hubble-like parameter. The action
is invariant under the transformations $\Phi\leftrightarrow\Phi$
and $H\leftrightarrow-H$. From this action, the Hamiltonian-like
constraint is derived:

\begin{equation}
\left.\frac{\delta I_{\mathrm{String}}^{2D}}{\delta n}\right|_{n=1}=0\qquad\Rightarrow\qquad\Phi^{\prime2}-\beta^{\prime2}-4\lambda^{2}=0.\label{eq:Hamiltonian constraint}
\end{equation}

\noindent To introduce the canonical momenta, we define:

\begin{equation}
\Pi_{\beta}=\left.\frac{\delta I_{\mathrm{String}}^{2D}}{\delta\beta^{\prime}}\right|_{n=1}=-\beta^{\prime}e^{-\Phi},\qquad\Pi_{\Phi}=\left.\frac{\delta I_{\mathrm{String}}^{2D}}{\delta\Phi^{\prime}}\right|_{n=1}=\Phi^{\prime}e^{-\Phi}.\label{eq:canonical momenta}
\end{equation}

\noindent The Hamiltonian is then:

\begin{equation}
H=\Pi_{\beta}^{2}-\Pi_{\Phi}^{2}+4\lambda^{2}e^{-2\Phi}.
\end{equation}

\noindent This Hamiltonian satisfies the momentum conservation condition:
$\left[\Pi_{\beta},H\right]=0$. 

\vspace{5mm}

To relate the wave functions to Schwarzschild spacetime and string
spacetime, we present the following classical solutions for $\Pi_{\beta}$
and $\Pi_{\Phi}$:

\subsubsection*{Classical solution for $\Pi_{\beta}$}

\noindent Using the solutions from equation (\ref{eq:configuration}),
we can verify the following:

\begin{eqnarray}
\rho<0, &  & \Pi_{\beta}=-\beta^{\prime}\left(-\rho\right)e^{-\Phi\left(-\rho\right)}=-c\lambda\equiv-k,\qquad\mathrm{String\;spacetime},\nonumber \\
\rho>0, &  & \Pi_{\beta}=-\beta^{\prime}\left(\rho\right)e^{-\Phi\left(\rho\right)}=-c\lambda\equiv-k,\qquad\mathrm{Schwarzschild\;spacetime},\label{eq:momentum conservation}
\end{eqnarray}
where $c\equiv\int d\rho\left(8\pi G_{2}\right)$ is a constant.

\subsubsection*{Classical solution for $\Pi_{\Phi}$}

Similarly, we can derive the solution for $\Pi_{\Phi}$:

\begin{eqnarray}
\rho<0, &  & \Pi_{\Phi}=\Phi^{\prime}\left(-\rho\right)e^{-\Phi\left(-\rho\right)}=c\lambda\cosh\left(2\rho\right),\qquad\mathrm{String\;spacetime},\nonumber \\
\rho>0, &  & \Pi_{\Phi}=\Phi^{\prime}\left(\rho\right)e^{-\Phi\left(\rho\right)}=-c\lambda\cosh\left(2\rho\right),\qquad\mathrm{Schwarzschild\;spacetime}.
\end{eqnarray}

\begin{figure}[h]
\begin{centering}
\includegraphics[scale=0.8]{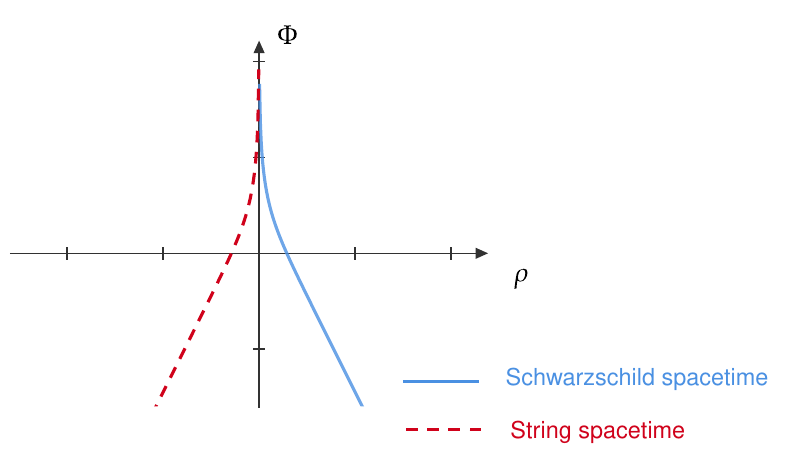}
\par\end{centering}
\caption{\label{fig:phirho} The relations between $\Phi\left(\rho\right)$
and $\rho$ for the classical solutions (\ref{eq:configuration})
describe how the dilaton field $\Phi\left(\rho\right)$ evolves as
a function of the spacelike coordinate $\rho$ in the reduced two-dimensional
near-horizon spacetime.}
\end{figure}

\noindent The relation between $\Phi\left(\rho\right)$ and $\rho$
is illustrated in Figure (\ref{fig:phirho}). 

\noindent At the strong coupling regime, where $\Phi\rightarrow+\infty$
corresponds to $\rho\rightarrow0$, we have the following limits:

\begin{eqnarray}
\rho<0, &  & \underset{\Phi\rightarrow+\infty}{\lim}\Pi_{\Phi}=c\lambda=k,\qquad\mathrm{String\;spacetime},\nonumber \\
\rho>0, &  & \underset{\Phi\rightarrow+\infty}{\lim}\Pi_{\Phi}=-c\lambda=-k,\qquad\mathrm{Schwarzschild\;spacetime}.\label{eq:large phi beta phi}
\end{eqnarray}

\noindent For $\Phi\rightarrow-\infty$ ($\rho\rightarrow\pm\infty$),
we have:

\begin{eqnarray}
\rho\rightarrow-\infty, &  & \underset{\Phi\rightarrow-\infty}{\lim}\Pi_{\Phi}=2\lambda e^{-\Phi}\equiv z,\qquad\mathrm{String\;spacetime},\nonumber \\
\rho\rightarrow+\infty, &  & \underset{\Phi\rightarrow-\infty}{\lim}\Pi_{\Phi}=-2\lambda e^{-\Phi}\equiv-z,\qquad\mathrm{Schwarzschild\;spacetime}.\label{eq:large phi beta phi -inf}
\end{eqnarray}

\vspace{5mm}

\noindent The corresponding Hamiltonian-like constraint from equation
(\ref{eq:Hamiltonian constraint}) is:

\begin{equation}
\Pi_{\beta}^{2}-\Pi_{\Phi}^{2}+4\lambda^{2}e^{-2\Phi}=0.
\end{equation}

\noindent Thus, we obtain the WDW equation for the action (\ref{eq:2D action})
through $\Pi_{\beta}\rightarrow i\partial_{\beta}$ and $\Pi_{\Phi}\rightarrow i\partial_{\Phi}$:

\begin{equation}
\left[\partial_{\Phi}^{2}-\partial_{\beta}^{2}+4\lambda^{2}e^{-2\Phi}\right]\Psi\left(\beta,\Phi\right)=0.\label{eq:singular WDW}
\end{equation}

\noindent To solve this equation, we employ the method of separation
of variables:

\begin{equation}
\Psi\left(\beta,\Phi\right)=\psi_{\beta}\psi_{\Phi}.\label{eq:singular WDW solution}
\end{equation}

\noindent Since

\begin{equation}
\Pi_{\beta}\Psi\left(\beta,\Phi\right)=i\partial_{\beta}\Psi\left(\beta,\Phi\right),
\end{equation}

\noindent and using equation (\ref{eq:momentum conservation}), the
wave function $\psi_{\beta}$ is given by:

\begin{eqnarray}
\rho<0, &  & \Pi_{\beta}\psi_{\beta}\left(\rho<0\right)=-k\psi_{\beta}\left(\rho<0\right)\rightarrow\psi_{\beta}\left(\rho<0\right)=e^{ik\beta},\qquad\mathrm{String\;spacetime},\nonumber \\
\rho>0, &  & \Pi_{\beta}\psi_{\beta}\left(\rho>0\right)=-k\psi_{\beta}\left(\rho>0\right)\rightarrow\psi_{\beta}\left(\rho>0\right)=e^{ik\beta},\qquad\mathrm{Schwarzschild\;spacetime}.
\end{eqnarray}

\noindent Therefore, we can fix the solution for $\psi_{\beta}$:

\begin{equation}
\psi_{\beta}=e^{ik\beta}.
\end{equation}

\noindent Since $\beta$ increases monotonically from $-\infty$ to
$+\infty$, it can be identified as a time-like coordinate in the
WDW equqaiton. Therefore, the tunneling will be triggered by the increasing
curvature. The complete solution thus becomes:

\begin{equation}
\Psi\left(\beta,\Phi\right)=\psi_{\Phi}e^{ik\beta}.
\end{equation}

\noindent Substituting this into the WDW equation (\ref{eq:singular WDW}),
we obtain:

\begin{equation}
\left[\partial_{\Phi}^{2}+k^{2}+4\lambda^{2}e^{-2\Phi}\right]\psi_{\Phi}=0.\label{eq:WDW equation}
\end{equation}

\noindent The general solution to this equation is:

\begin{equation}
\psi_{\Phi}=c_{1}\Gamma\left(1-\nu\right)J_{-\nu}\left(z\right)+c_{2}\Gamma\left(1+\nu\right)J_{+\nu}\left(z\right).
\end{equation}

\noindent where $\Gamma$ is the Euler gamma function, $J_{\pm\nu}\left(z\right)$
are Bessel functions, $\nu=ik$, and $z=2\lambda e^{-\Phi}$. Before
proceeding further, it is essential to clarify the physical significance
of this linear combination of Bessel functions. As is well known:

\begin{equation}
\underset{\Phi\rightarrow+\infty}{\lim}J_{\pm ik}\left(2\lambda e^{-\Phi}\right)\sim e^{\mp ik\Phi},
\end{equation}

\noindent Based on the relations (\ref{eq:large phi beta phi}) and
$\Pi_{\Phi}\Psi\left(\beta,\Phi\right)=i\partial_{\Phi}\Psi\left(\beta,\Phi\right)$,
we can identify the wave function regions:

\begin{equation}
\psi_{\Phi}\left(\rho<0\right)=c_{2}\Gamma\left(1+\nu\right)J_{+\nu}\left(z\right),\qquad\psi_{\Phi}\left(\rho>0\right)=c_{1}\Gamma\left(1-\nu\right)J_{-\nu}\left(z\right).
\end{equation}

\noindent As $\Phi\rightarrow+\infty$, the potential in the WDW equation
(\ref{eq:WDW equation}) vanishes, and the plane-wave solution becomes:

\begin{equation}
\psi_{\Phi\rightarrow+\infty}=\psi_{\Phi\rightarrow+\infty}\left(\rho<0\right)+\psi_{\Phi\rightarrow+\infty}\left(\rho>0\right)\sim e^{-ik\Phi}+e^{ik\Phi},
\end{equation}

\noindent Thus, the complete wave function in the strong coupling
region can be written as a superposition of right-moving ($\rho<0$)
and left-moving ($\rho>0$) waves:

\begin{equation}
\Psi\left(\beta,\Phi\rightarrow+\infty\right)=\psi_{\Phi\rightarrow+\infty}\left(\rho<0\right)e^{ik\beta}+\psi_{\Phi\rightarrow+\infty}\left(\rho>0\right)e^{ik\beta}.
\end{equation}

\noindent Before imposing the boundary condition, we first clarify
the value range of $\beta$. Referring to the classical solutions
(\ref{eq:configuration}), we find:

\begin{eqnarray}
\rho<0, &  & \beta\left(-\rho\right)=\ln\left(\coth\left(-\lambda\rho\right)\right)>0,\qquad\mathrm{String\;spacetime},\nonumber \\
\rho>0, &  & \beta\left(\rho\right)=\ln\left(\tanh\left(\lambda\rho\right)\right)<0,\qquad\mathrm{Schwarzschild\;spacetime}.\label{eq:beta range}
\end{eqnarray}

\noindent Now, consider the initial wave incoming from the low-energy
regime, where $\beta<0$ and $\Phi\rightarrow-\infty$. This wave
corresponds to the Schwarzschild spacetime, as determined by (\ref{eq:beta range}).
In this paper, we are only interested in a specific boundary condition---namely,
the tunneling boundary condition---such that only the right-moving
wave, $\psi_{\Phi}\left(\rho>0\right)$, evolves toward the singularity
at $\beta>0$ and $\Phi\rightarrow+\infty$ (strong coupling regime)
\cite{Vilenkin:1986cy,Vilenkin:1987kf}. It is worth noting that although
the divergence of the dilaton indicates a breakdown of the classical
description near the horizon, our methodology (taking a large $D$
limit) considerably suppresses fluctuations in the transverse sphere.
Hence, the dominant dynamics remain effectively two-dimensional, allowing
the semiclassical approximation to capture the essential tunneling
physics. In the future work, it is possible to study the higher\nobreakdash-order
corrections to this process \cite{Ying:2024wrv}. Under this boundary
condition, the specific solution is given by:

\begin{equation}
\Psi\left(\beta,\Phi\right)=\psi_{\Phi}\left(\rho>0\right)e^{ik\beta}=c_{1}\Gamma\left(1-\nu\right)J_{-\nu}\left(z\right)e^{ik\beta}.
\end{equation}

\noindent On the other hand, in the regime $\Phi\rightarrow-\infty$
(or equivalently $z\rightarrow+\infty$), the solution can be expanded
as:

\begin{equation}
\underset{\Phi\rightarrow-\infty}{\lim}\Psi\left(\beta,\Phi\right)=\frac{c_{1}e^{ik\beta}}{\sqrt{2\pi z}}\left[\exp\left(i\left(z-\frac{\pi}{4}\right)\right)\exp\left(-\frac{k\pi}{2}\right)+\exp\left(-i\left(z-\frac{\pi}{4}\right)\right)\exp\left(\frac{k\pi}{2}\right)\right].
\end{equation}

\noindent Based on (\ref{eq:large phi beta phi -inf}), we identify
the wave functions in the following limits:

\begin{eqnarray}
\rho\rightarrow+\infty, &  & \Psi_{-}\left(\beta,\Phi\right)=\frac{c_{1}e^{ik\beta}}{\sqrt{2\pi z}}\exp\left(i\left(z-\frac{\pi}{4}\right)\right)\exp\left(-\frac{k\pi}{2}\right),\qquad\mathrm{String\;spacetime}\nonumber \\
\rho\rightarrow-\infty, &  & \Psi_{+}\left(\beta,\Phi\right)=\frac{c_{1}e^{ik\beta}}{\sqrt{2\pi z}}\exp\left(-i\left(z-\frac{\pi}{4}\right)\right)\exp\left(\frac{k\pi}{2}\right),\qquad\mathrm{Schwarzschild\;spacetime},
\end{eqnarray}

\noindent through

\begin{equation}
\underset{\Phi\rightarrow-\infty}{\lim}\Pi_{\Phi}\Psi_{\pm}\left(\beta,\Phi\right)=\underset{\Phi\rightarrow-\infty}{\lim}\mp z\Psi\left(\beta,\Phi\right).
\end{equation}

\noindent Consequently, we consider the incoming wave $\Psi_{+}\left(\beta,\Phi\rightarrow-\infty\right)$,
representing the Schwarzschild black hole, originating from the low-energy
limit $\beta<0$ and $\Phi\rightarrow-\infty$. This wave partially
transmits to the singularity at $\beta>0$ and $\Phi\rightarrow+\infty$,
while the remaining portion, $\Psi_{-}\left(\beta,\Phi\rightarrow-\infty\right)$
reflects to the string black hole. Finally, the transition can be
understood as a reflection of the wave function in the minisuperspace
$\left(\beta,\Phi\right)$. The transition probability is then determined
by the reflection coefficient:

\begin{equation}
P=\frac{\left|\Psi_{-}\left(\beta,\Phi\rightarrow-\infty\right)\right|^{2}}{\left|\Psi_{+}\left(\beta,\Phi\rightarrow-\infty\right)\right|^{2}}=\exp\left(-2k\pi\right).
\end{equation}

\noindent This result implies that although the corresponding transition
being classically forbidden, this quantum process has a nonzero probability.
The similar result in string cosmology can be found in the refs. \cite{Gasperini:1996fn,Gasperini:2007zz}.
Remarkably, this result is consistent with predictions from loop quantum
gravity. As the black hole approaches the final stages of its evaporation,
the probability of tunneling into a white hole is no longer non-perturbatively
suppressed \cite{Bianchi:2018mml}. Moreover, in both cases, the tunneling
occurs near the horizon, $r\sim r_{0}$. Finally, this result not
only provides the transition probability between a Schwarzschild black
hole and a string black hole but also serves as a counterexample to
the weak cosmic censorship hypothesis, suggesting that a black hole
with an event horizon may transition to a black hole with a naked
singularity.

\section{Conclusion}

In this paper, we investigated the large $D$ limit of T-dual Schwarzschild
and string black holes. In this limit, the near-horizon geometries
reduced to two-dimensional black string solutions. These two geometries
were T-dual to each other, shared the same low-energy effective action,
and described different regions of the two-dimensional target space.
Due to this property, we derived the corresponding WDW equation for
this two-dimensional background, which enabled us to study the transition
between the two T-dual geometries. The transition probability was
also computed. Our results demonstrated that the classically forbidden
transition between Schwarzschild and string black holes can be realized
in the large $D$ limit through quantum dynamics.

In future work, we aim to explore the following topics:
\begin{itemize}
\item After deriving the transition probability between the near-horizon
geometries of large $D$ Schwarzschild and string black holes, it
will be valuable to investigate how this transition extends to fundamental
strings.
\item We are interested in exploring potential observable consequences of
this result. Specifically, since Schwarzschild black holes and naked
string black holes have different photon spheres, a transition between
them should manifest in observable changes in the photon sphere.
\item The weak cosmic censorship conjecture is also a topic worth re-examining
in this new framework.
\end{itemize}
\vspace{5mm}

\noindent {\bf Acknowledgements} 
This work was supported by NSFC Grants No.12105031, No.12347101.


\begin{thebibliography}{99}

\bibitem{Bowick:1985af} M.~J.~Bowick, L.~Smolin and L.~C.~R.~Wijewardhana, ``Role of String Excitations in the Last Stages of Black Hole Evaporation,'' Phys. Rev. Lett. \textbf{56}, 424 (1986) doi:10.1103/PhysRevLett.56.424 

\bibitem{Susskind:1993ws} L.~Susskind, ``Some speculations about black hole entropy in string theory,'' [arXiv:hep-th/9309145 [hep-th]]. 

\bibitem{Horowitz:1996nw} G.~T.~Horowitz and J.~Polchinski, ``A Correspondence principle for black holes and strings,'' Phys. Rev. D \textbf{55}, 6189-6197 (1997) doi:10.1103/PhysRevD.55.6189 [arXiv:hep-th/9612146 [hep-th]]. 

\bibitem{Horowitz:1997jc} G.~T.~Horowitz and J.~Polchinski, ``Selfgravitating fundamental strings,'' Phys. Rev. D \textbf{57}, 2557-2563 (1998) doi:10.1103/PhysRevD.57.2557 [arXiv:hep-th/9707170 [hep-th]]. 



\bibitem{Chen:2021emg} Y.~Chen and J.~Maldacena, ``String scale black holes at large D,'' JHEP \textbf{01}, 095 (2022) doi:10.1007/JHEP01(2022)095 [arXiv:2106.02169 [hep-th]]. 


\bibitem{Chen:2021dsw} Y.~Chen, J.~Maldacena and E.~Witten, ``On the black hole/string transition,'' JHEP \textbf{01}, 103 (2023) doi:10.1007/JHEP01(2023)103 [arXiv:2109.08563 [hep-th]]. 

\bibitem{Ceplak:2023afb} N.~\v{C}eplak, R.~Emparan, A.~Puhm and M.~Toma\v{s}evi\'c, ``The correspondence between rotating black holes and fundamental strings,'' JHEP \textbf{11}, 226 (2023) doi:10.1007/JHEP11(2023)226 [arXiv:2307.03573 [hep-th]]. 

\bibitem{Chu:2024ggi} 
J.~Chu, ``From Black Strings to Fundamental Strings: Non-uniformity and Phase Transitions,'' [arXiv:2410.23597 [hep-th]]. 







\bibitem{Tseytlin:1991wr} A.~A.~Tseytlin, ``Duality and dilaton,'' Mod. Phys. Lett. A \textbf{6}, 1721-1732 (1991) doi:10.1142/S021773239100186X 

\bibitem{Ginsparg:1992af} P.~H.~Ginsparg and F.~Quevedo, ``Strings on curved space-times: Black holes, torsion, and duality,'' Nucl. Phys. B \textbf{385}, 527-557 (1992) doi:10.1016/0550-3213(92)90057-I [arXiv:hep-th/9202092 [hep-th]]. 

\bibitem{Kar:1998rv} S.~Kar, ``Naked singularities in low-energy, effective string theory,'' Class. Quant. Grav. \textbf{16}, 101-115 (1999) doi:10.1088/0264-9381/16/1/008 [arXiv:hep-th/9804039 [hep-th]]. 


\bibitem{Exirifard:2004ey} G.~Exirifard and M.~O'Loughlin, ``Two and three loop alpha-prime corrections to T-duality: Kasner and Schwarzschild,'' JHEP \textbf{12}, 023 (2004) doi:10.1088/1126-6708/2004/12/023 [arXiv:hep-th/0408200 [hep-th]]. 



\bibitem{Emparan:2013xia} R.~Emparan, D.~Grumiller and K.~Tanabe, ``Large-D gravity and low-D strings,'' Phys. Rev. Lett. \textbf{110}, no.25, 251102 (2013) doi:10.1103/PhysRevLett.110.251102 [arXiv:1303.1995 [hep-th]]. 

\bibitem{Ying:2024wrv} S.~Ying, ``Large D gravity and low D string via $\alpha^{\prime}$ corrections,'' JHEP \textbf{09}, 156 (2024) doi:10.1007/JHEP09(2024)156 [arXiv:2407.18179 [hep-th]]. 


\bibitem{Gasperini:1991ak} M.~Gasperini and G.~Veneziano, ``O(d,d) covariant string cosmology,'' Phys. Lett. B \textbf{277}, 256-264 (1992) doi:10.1016/0370-2693(92)90744-O [arXiv:hep-th/9112044 [hep-th]]. 





\bibitem{Gasperini:1996fn} M.~Gasperini and G.~Veneziano, ``Birth of the universe as quantum scattering in string cosmology,'' Gen. Rel. Grav. \textbf{28}, 1301-1307 (1996) doi:10.1007/BF02109522 [arXiv:hep-th/9602096 [hep-th]]. 

\bibitem{Gasperini:1996np} M.~Gasperini, J.~Maharana and G.~Veneziano, ``Graceful exit in quantum string cosmology,'' Nucl. Phys. B \textbf{472}, 349-360 (1996) doi:10.1016/0550-3213(96)00201-5 [arXiv:hep-th/9602087 [hep-th]]. 


\bibitem{Gasperini:1997uh} M.~Gasperini, ``Low-energy quantum string cosmology,'' Int. J. Mod. Phys. A \textbf{13}, 4779-4786 (1998) doi:10.1142/S0217751X98002250 [arXiv:hep-th/9706049 [hep-th]]. 

\bibitem{Gasperini:2007zz} M.~Gasperini, ``Elements of string cosmology,'' Cambridge University Press, 2007, ISBN 978-0-511-33229-6, 978-0-521-18798-5, 978-0-521-86875-4 


\bibitem{Gasperini:2021eri} M.~Gasperini, ``Quantum string cosmology,'' Universe \textbf{7}, no.1, 14 (2021) doi:10.3390/universe7010014 [arXiv:2101.01070 [gr-qc]]. 







\bibitem{Veneziano:1991ek} G.~Veneziano, ``Scale factor duality for classical and quantum strings,'' Phys. Lett. B \textbf{265}, 287-294 (1991) doi:10.1016/0370-2693(91)90055-U 

\bibitem{Sen:1991zi} A.~Sen, ``O(d) x O(d) symmetry of the space of cosmological solutions in string theory, scale factor duality and two-dimensional black holes,'' Phys. Lett. B \textbf{271}, 295-300 (1991) doi:10.1016/0370-2693(91)90090-D 



\bibitem{Vilenkin:1986cy} A.~Vilenkin, ``Boundary Conditions in Quantum Cosmology,'' Phys. Rev. D \textbf{33}, 3560 (1986) doi:10.1103/PhysRevD.33.3560 

\bibitem{Vilenkin:1987kf} A.~Vilenkin, ``Quantum Cosmology and the Initial State of the Universe,'' Phys. Rev. D \textbf{37}, 888 (1988) doi:10.1103/PhysRevD.37.888 



\bibitem{Bianchi:2018mml} 
E.~Bianchi, M.~Christodoulou, F.~D'Ambrosio, H.~M.~Haggard and C.~Rovelli, ``White Holes as Remnants: A Surprising Scenario for the End of a Black Hole,'' Class. Quant. Grav. \textbf{35}, no.22, 225003 (2018) doi:10.1088/1361-6382/aae550 [arXiv:1802.04264 [gr-qc]]. 

 
\end{thebibliography}
\end{document}